\documentclass[aps,prb,twocolumn,superscriptaddress,showpacs]{revtex4-1}

\usepackage{graphicx} 
\usepackage{color}
\usepackage{float}

\begin{document}


\title{Magnetic Neutron Scattering of Thermally Quenched K-Co-Fe Prussian Blue Analogue Photomagnet}


\author{Daniel M. Pajerowski}
\affiliation{NIST Center for Neutron Research, Gaithersburg, MD 20899-6012, USA}
\affiliation{Department of Physics and National High Magnetic Field Laboratory, University of Florida, Gainesville, FL 32611-8440, USA}

\author{V. Ovidiu Garlea}
\affiliation{Quantum Condensed Matter Division, Oak Ridge National Laboratory, Oak Ridge, TN 37831-6393, USA}

\author{Elisabeth S. Knowles}
\affiliation{Department of Physics and National High Magnetic Field Laboratory, University of Florida, Gainesville, FL 32611-8440, USA}

\author{Matthew J. Andrus}
\affiliation{Department of Chemistry, University of Florida, Gainesville, FL 32611-7200, USA}

\author{Matthieu F. Dumont}
\affiliation{Department of  Physics and National High Magnetic Field Laboratory, University of Florida, Gainesville, FL 32611-8440, USA}
\affiliation{Department of Chemistry, University of Florida, Gainesville, FL 32611-7200, USA}

\author{Yitzi M. Calm}
\affiliation{Department of Physics and National High Magnetic Field Laboratory, University of Florida, Gainesville, FL 32611-8440, USA}

\author{Stephen E. Nagler}
\affiliation{Quantum Condensed Matter Division, Oak Ridge National Laboratory, Oak Ridge, TN 37831-6393, USA}

\author{Xin Tong}
\affiliation{Instrument and Source Design Division, Oak Ridge National Laboratory, Oak Ridge, TN 37831-6393, USA}

\author{Daniel R. Talham}
\affiliation{Department of Chemistry, University of Florida, Gainesville, FL 32611-7200, USA}

\author{Mark W. Meisel}
\affiliation{Department of Physics and National High Magnetic Field Laboratory, University of Florida, Gainesville, FL 32611-8440, USA}

\date{\today}

\begin{abstract}
Magnetic order in the thermally quenched photomagnetic Prussian blue analogue coordination polymer 
K$_{0.27}$Co[Fe(CN)$_6$]$_{0.73}$[{D$_2$O}$_6$]$_{0.27}\cdot$1.42D$_2$O has been studied down to 4 K 
with unpolarized and polarized neutron powder diffraction as a function of applied magnetic field.  
Analysis of the data allows the onsite coherent magnetization of the Co and Fe spins to be 
established.  Specifically, magnetic fields of 1~T and 4~T 
induce moments parallel to the applied field, and the sample behaves as a ferromagnet with a 
wandering axis.
\end{abstract}

\pacs{75.50.Xx, 75.25.-j, 75.30.Gw, 75.50.LK}

\maketitle



\section{Introduction}

\begin{figure}[b]
	\includegraphics[width=65mm]{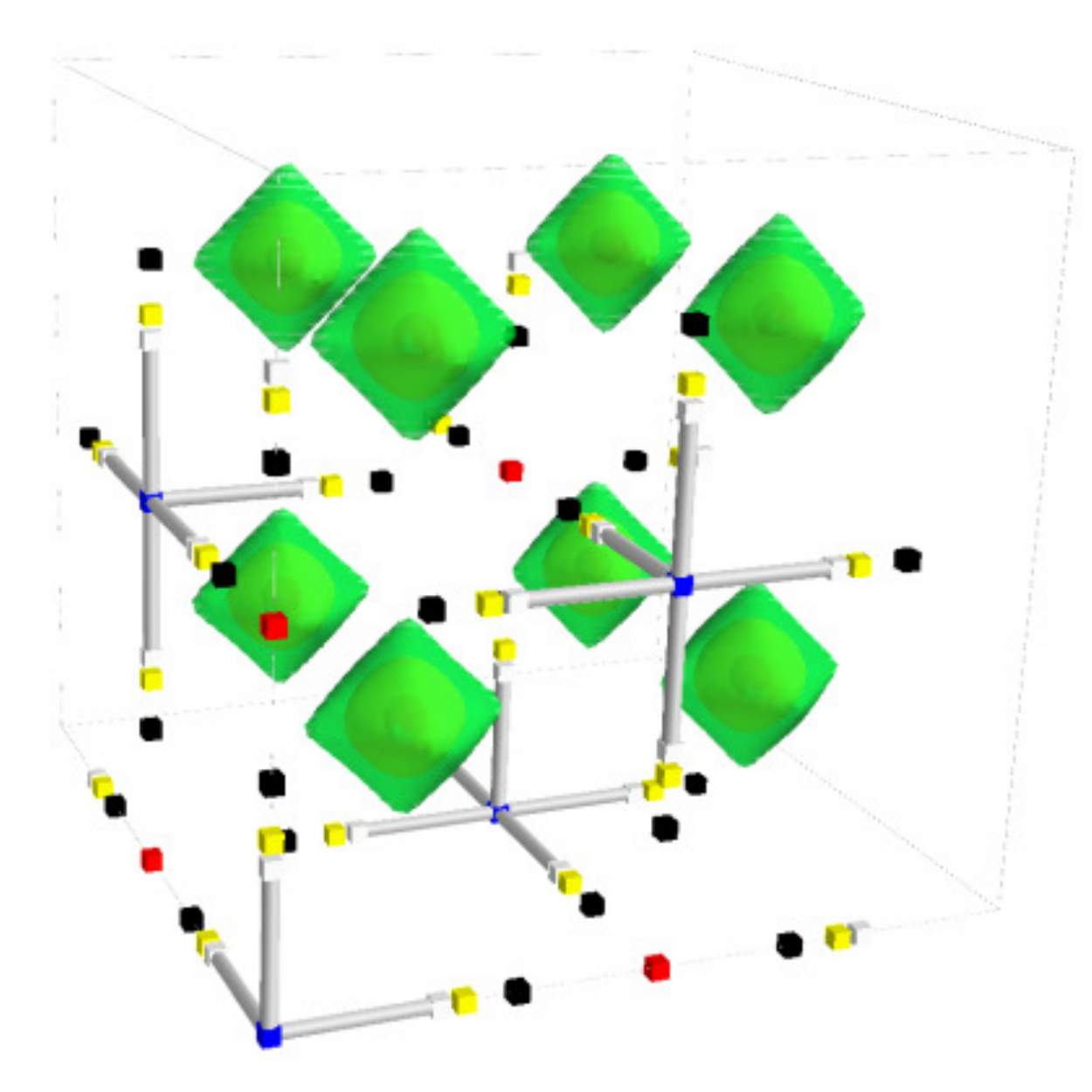}
	\caption{(Color online) Co-Fe unit cell.  Crystallographic positions of atoms within the unit cell are illustrated with cubes for K (cyan), Co (red), Fe (blue), C (white), N (black), and coordinated D$_2$O, O positions (yellow), while the interstitial D$_2$O density is displayed using contoured isosurfaces (green) and Fe-C bonds are displayed as tubes (white).  Details of structure determination are presented in Section III.}
	\label{fig:KCoFeFig1}
\end{figure}

Manipulating magnetization with photons is now a major research focus because it may yield materials capable of dense information storage.  An epitomic example of a photomagnetic coordination polymer is potassium cobalt hexacyanoferrate, K$_{\alpha}$Co[Fe(CN)$_6$]$_{\beta}\cdot$nH$_{2}$O (from now on referred to as Co-Fe, with the crystal structure shown in Fig.~\ref{fig:KCoFeFig1}), which displays magnetic order and an optical charge transfer induced spin transition (CTIST).\cite{Sato1996}  The details of the magnetism in Co-Fe have been investigated with bulk probes such as magnetization,\cite{Sato1999} and AC-susceptibility,\cite{Pejakovic2002} as well as atomic level probes such as X-ray magnetic circular dichroism (XMCD),\cite{Champion2001} and muon spin relaxation ($\mu$-SR).\cite{Salman2006}  However, we utilize neutron scattering because it is capable of extracting the magnetic structure, including the length and direction of the magnetic moments associated with different crystallographic positions.

Neutron scattering research has been important in understanding the structure of materials similar to Co-Fe.  
For example, neutron diffraction has been used to elucidate the location of water molecules, 
to identify the long-range magnetic order, and to explore the spin delocalizetion in Prussian 
blue.\cite{Buser1977,Herren1980,Day1980}  
Later work used similar techniques to investigate hydrogen adsorption in Cu$_{3}$[Co(CN)$_{6}$]$_{2}$, along with vibrational spectroscopy,\cite{Hartman2006} and neutron vibrational spectroscopy was also measured in Zn$_{3}$[Fe(CN)$_{6}$]$_{2}$.\cite{Adak2010}  Likewise, magnetic structure determination with neutrons was used to explore negative magnetization in Cu$_{0.73}$Mn$_{0.77}$[Fe(CN)$_{6}]\cdot$nH$_{2}$O,\cite{Kumar2008} and to extract on-site moments in Berlin green\cite{Kumar2005,Kumar2004} and in (Ni$_x$Mn$_{1-x}$)$_3$[Cr(CN)$_6$]$_2$ molecule-based magnets.\cite{Mihalik2010}

To this end, we have performed neutron powder diffraction (NPD) on deuterated Co-Fe samples in magnetic states resulting from thermal quenching.  Briefly, at room temperature, the photomagnetic Co-Fe with optimal iron vacancies is paramagnetic, with a transition to the diamagnetic low-spin state when cooling below nominally 200~K.  It is below 100~K that applied light may convert molecules from diamagnetic to paramagnetic and back, and below around 20~K where this effect is most striking due to the large susceptibility of the magnetically ordered state.  However, a magnetically ordered state may also be achieved at low temperatures by thermally quenching,\cite{Park2007,Chong2011} where the paramagnetic 300~K state is cooled so quickly to below 100~K that it does not relax to the diamagnetic ground state.  It is this magnetic, thermally quenched state that we study with NPD as a function of magnetic field, while complementary magnetization, transmission electron microscopy (TEM), Fourier transform infrared spectroscopy (FT-IR), and elemental analysis have also been performed on the sample.  We find that Co-Fe possesses a correlated spin glass ground state that is driven via magnetic field to behave as a ferromagnet with a wandering axis.

\section{Experimental Details\cite{NISTdisclaimer}}

\subsection{Synthesis}
To begin preparation of K$_{\alpha}$Co[Fe(CN)$_6$]$_{\beta}\cdot$nD$_{2}$O powder, a 75 mL solution of 0.1 mol/L KNO$_{3}$ in D$_{2}$O was added to a 75 mL solution of 20$\times$10$^{-3}$ mol/L K$_{3}$[Fe(CN)$_{6}$] in D$_{2}$O, and stirred for ten minutes.  While continuing to stir, a 300 mL solution of 5$\times$10$^{-3}$ mol/L CoCl$_{2}$ in D$_{2}$O was added drop-wise over the course of two hours.  Stirring of the final solution was allowed to continue two additional hours subsequent to complete mixing.  Next, the precipitate was collected by centrifugation at 2000 rpm (210 rad/s) for 10 min (600 s) and dried under vacuum.  This procedure was repeated 14 times until 4.37 g of powder was collected.  Potassium ferricyanide, anhydrous cobalt chloride, and potassium nitrate were all purchased from Sigma-Aldrich. To remove water, the potassium nitrate was heated in an oven to 110 $^\circ$C (383~K) for 4 h before use.  All other reagents and chemicals were used without further purification.  Deuterium oxide was purchased from Cambridge Isotope Laboratories, Inc.

\subsection{Instrumentation}
Neutron powder-diffraction experiments were conducted using the HB2A diffractometer at the High Flux Isotope Reactor,\cite{Garlea2010} using the Ge[113] monochromator with $\lambda$  = 2.41 $\mathrm{\AA}$ (0.241 nm).  Sample environment on HB2A utilized an Oxford 5 T, vertical-field magnet with helium cryogenics.  Neutron polarization was achieved with a $^3$He cell that produced 79$\%$ polarization at the beginning of the experiment and decayed to 63$\%$ polarization after 20 hours at the end of the experiment, to give an average polarization of 71$\%$ for both up and down polarization measurements, and we did not perform polarization analysis after the sample but instead followed established methods for powder diffraction with polarized neutrons.\cite{Lelievre2010,Wills2005} The flipping difference spectra were obtained by subtracting the diffraction data, measured with the incident neutron polarization parallel to the applied field and magnetization, from the data recorded with the incident polarization antiparallel to the field. Magnetic measurements were performed using a Quantum Design MPMS XL superconducting quantum interference device (SQUID) magnetometer.  Infrared spectra were recorded on a Thermo Scientific Nicolet 6700 spectrometer.  Energy dispersive X-ray spectroscopy (EDS) and TEM were conducted on a JEOL 2010F super probe by the Major Analytical Instrumentation Center at the University of Florida (UF).  The UF Spectroscopic Services Laboratory performed combustion analysis.

\subsection{Analysis Preparations}
For NPD, 4.37 g of powder were mounted in a cylindrical aluminum can.  Thermal quenching to trap the magnetic state was achieved by filling the cryostat bath with liquid helium and directly inserting a sample stick from ambient temperature.  To avoid hydrogen impurities, the powder was wetted with deuterium oxide, and to avoid sample movement in magnetic fields, an aluminum plug was inserted above the sample.  To measure magnetization, samples heavier than 10 mg were mounted in gelcaps and held in plastic straws.  Thermal quenching in the SQUID was achieved by equilibrating the cryostat to 100~K, and directly inserting the sample stick from ambient temperature.  For measurements in 10~mT, samples are cooled through the ordering temperature in 10~mT, and for 1~T and 4~T measurements, there is no observed thermal hysteresis.  For FT-IR, less than 1 mg amounts of sample were suspended in an acetone solution and deposited on KBr salt plates and allowed to dry.  For EDS and TEM, acetone suspensions of the powder were deposited onto 400 mesh copper grids with an ultrathin carbon film on a holey carbon support obtained from Ted Pella, Inc.

\subsection{Diffraction analysis scheme}
Intensities were fit to the standard powder diffraction equation with a correction for absorption,
\begin{eqnarray}
I(\theta)&=&A_0\frac{m_{hkl}  |F (hkl )|^2}{\sin\theta \, \sin2\theta} \;\eta (\theta)~~~,
\end{eqnarray}
with
\begin{eqnarray}
\eta (\theta) &=&  e ^{-(1.713 - 0.037 \sin^2\theta )\mu R + (0.093 + 0.375  \sin^2 \theta  ) \mu ^2  R ^2  },
\end{eqnarray}
where $A_0$ is an overall scale factor, $m_{hkl}$ is the multiplicity of the scattering vector, 
$F$ is the structure factor, $\theta$ is the scattering angle, $\mu$ is the linear attenuation coefficient, 
and $R$ is the radius of the sample cylinder.
For our sample and experimental arrangement, $\mu R = 0.17$, which has little effect on 
the observed intensities aside from scale.  The structure factor has nuclear ($F_N$) and magnetic ($F_M$) 
contributions, and for unpolarized neutrons 
\begin{equation}
|F|^2~=~|F_N+F_M|^2~=~|F_N|^2 + |F_M|^2 ~~~.
\end{equation}
On the other hand, $F_N$ and $F_M$ can coherently interfere for polarized neutrons such that, for moments co-linear with $P$, 
\begin{equation}
|F|^2~=~|F_N+F_M|^2~=~|F_N|^2 + |F_M|^2 \pm 2PF_NF_M~~~,
\end{equation}
where $P$ is the neutron polarization fraction and the sign of the final term depends upon up or down neutron polarization.\cite{Schweizer2006}  For nuclear scattering,
\begin{equation}
F_N(hkl)~=~\sum_{j}{n_j b_j  e^{iG\cdot d_j} e^{-W_j}}~~~,
\end{equation}
where the sum is over all atoms in the unit cell, $n$ is related to the average occupancy, $b$ is the coherent nuclear scattering length, $G$ is the $hkl$ dependent reciprocal lattice vector, $d$ is the direct space atomic position, and $W = BQ^2/16 \pi ^2$ is the Debye-Waller factor.  For magnetic scattering, all coherent scattering is modeled to be along the applied field, which is perpendicular to the scattering plane, so that
\begin{equation}
F_M(hkl )~=~\frac{\gamma r_0}{2} \sum_{j} {m_j(Q) e^{iG \cdot d_j  } e ^{-W_j} }~~~,
\end{equation}
where $\frac{\gamma r_0}{2}~=~2.695$~fm, and the magnetization can be written as
\begin{eqnarray}
m_j(Q)&=& \langle L_z \rangle _j  f_{L,j}(Q )~+~2 \langle S_z \rangle _j  f_{S,j} (Q )  \nonumber\\
&=&g_{J,j}  \langle J_z \rangle _j f_{J,j}(Q )  \nonumber\\
&=&\langle J_z \rangle _j  (g_{L,j} f_{L,j}(Q )~+~g_{S,j}  f_{S,j}(Q ))~~~,
\end{eqnarray}
where $\langle J_z \rangle$ is the average total angular momentum, $\langle L_z \rangle$ is the average 
orbital angular momentum, $\langle S_z \rangle$ is the average spin angular momentum, $f_J(Q)$ is the 
magnetic form factor for the total angular momentum, $f_L(Q)$ is the magnetic form factor for the 
orbital angular momentum, $f_S(Q)$ is the magnetic form factor for the spin angular momentum, and 
$g_J$, $g_L$ and $g_S$ may be determined by Wigner's formula.\cite{Sakurai1994}  The tabulated form 
factor values within the dipole approximation are used for the spin and orbital form 
factors.\cite{Clementi1974}  Squared differences between observed and calculated intensities were 
minimized using a Nelder-Mead simplex algorithm.  Open-source Python 2.7 libraries were utilized 
to aid in plotting routines, matplotlib 1.0.1 and Mayavi2, and computation, NumPy 1.6.1 and SciPy 0.7.2.  
Reported uncertainties of fit parameters are the square root of the diagonal terms in the covariance 
matrix multiplied by the standard deviation of the residuals.

\begin{figure}[t]
	\includegraphics[width=87mm]{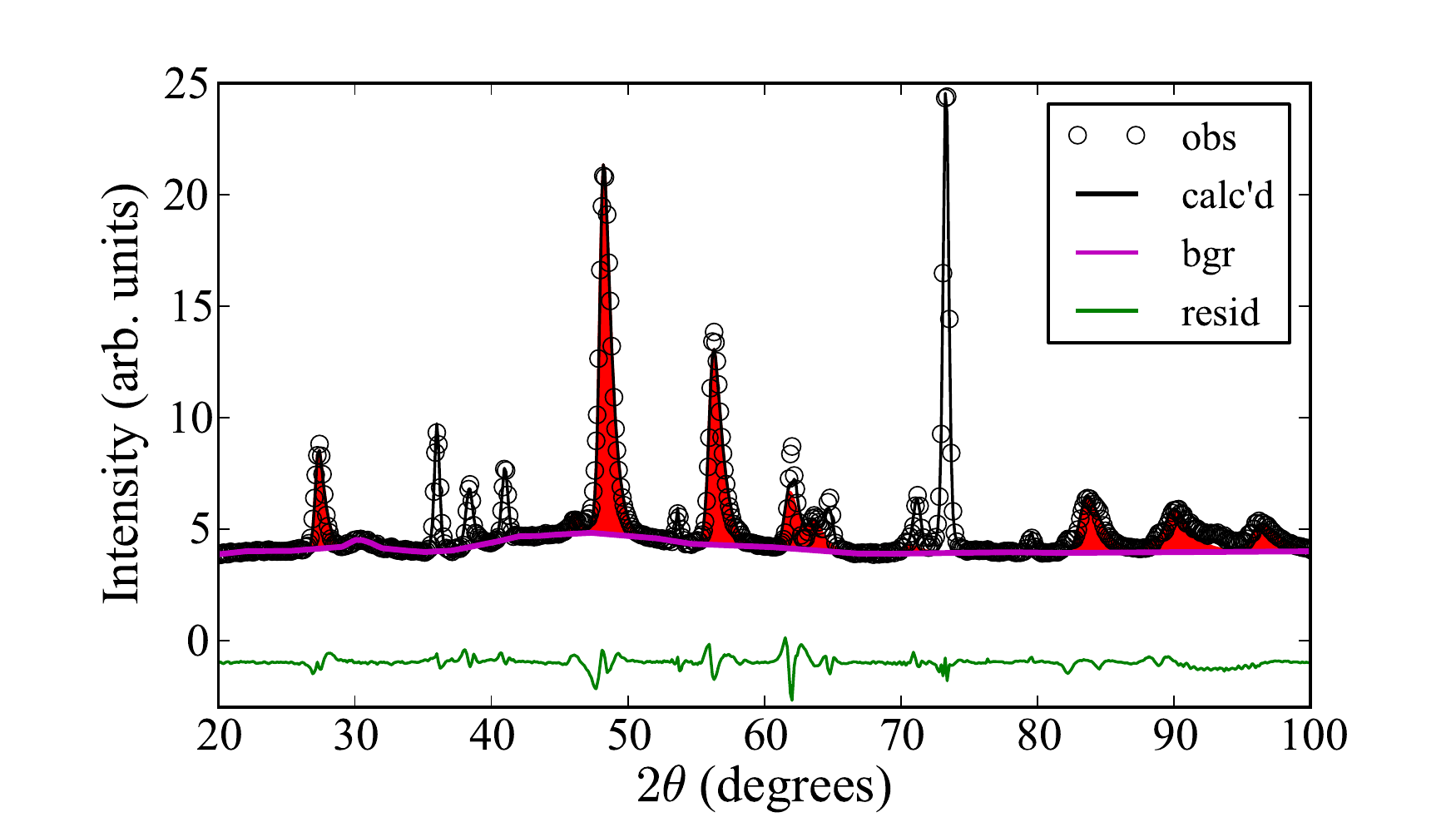}
	\caption{(Color online) Neutron powder diffraction of Co-Fe at $T$~=~40~K.  Observed scattering is shown as open circles (obs), a full fit including sample mount contributions is shown as a black line (calc'd), the diffuse background is illustrated with a magenta line (bgr), and the residuals of the fit are shown below the zero line with a green line (resid).  The signal due to Co-Fe is emphasized with a red filling.  Experimental uncertainties derived from counting statistics are smaller than the plotting symbols.}
	\label{fig:KCoFeFig2}
\end{figure}

\begin{table} [h]
\caption{Atomic coordinates and occupancies for Co-Fe at $T$~=~40~K.\label{table1}}
\begin{ruledtabular}
\begin{tabular}{ c c c c c c }
atom & position & $n$ & x & y & z\\
\hline
Co & 4a & 1 & 0.5 & 0.5 & 0.5\\
Fe & 4b & 0.73 & 0 & 0 & 0\\
C & 24e & 0.73 & 0.212 & 0 & 0\\
N & 24e & 0.73 & 0.313 & 0 & 0\\
K & 8c & 0.135 & 0.25 & 0.25 & 0.25\\
O & 24e & 0.27 & 0.243 & 0 & 0\\
D & 96k & 0.135 & 0.303 & 0.060 & 0.060\\
\end{tabular}
\end{ruledtabular}

\end{table}

\begin{figure*} [ht!]
	\includegraphics[width=170mm]{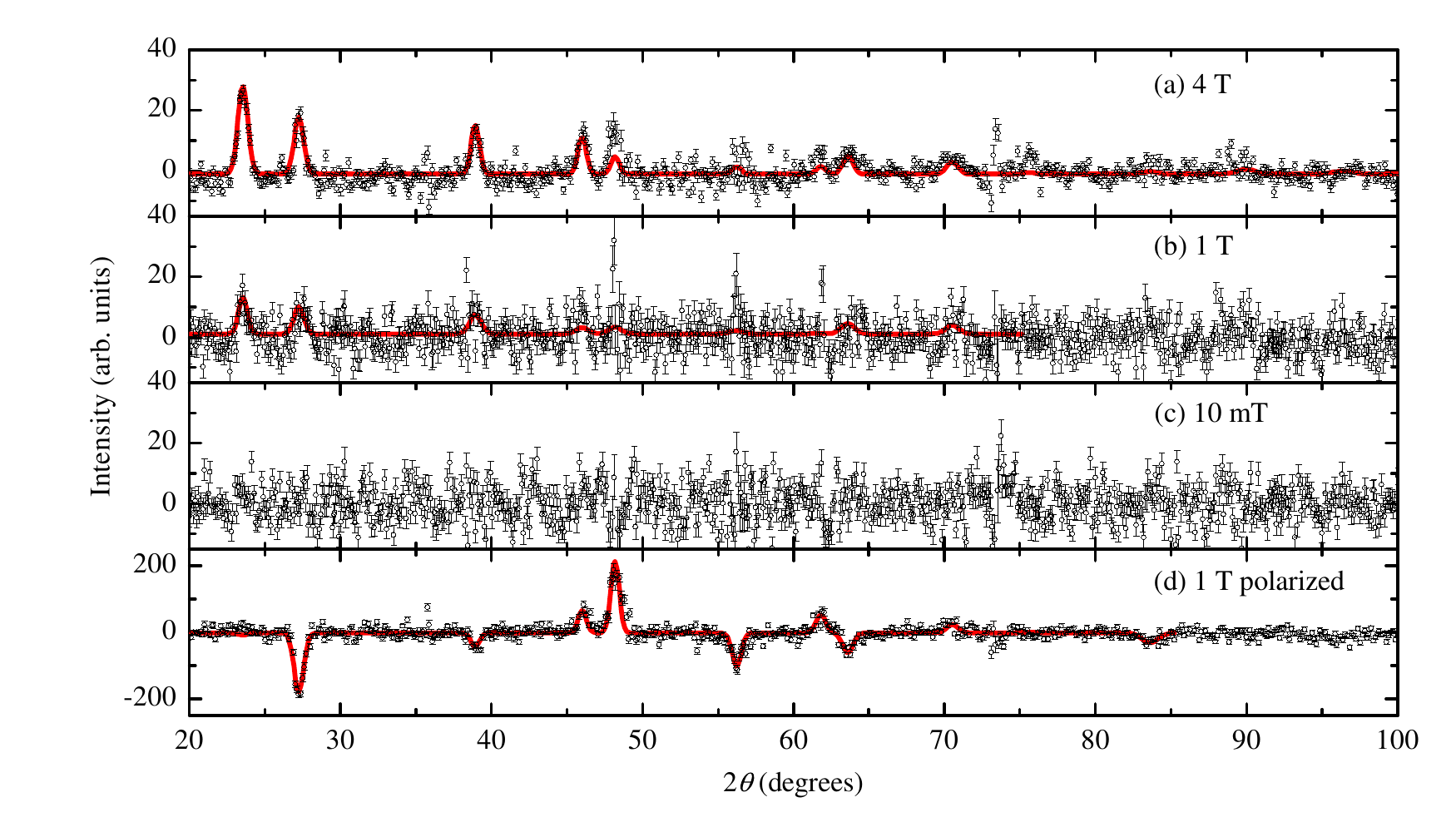}
	\caption{(Color online) Magnetic neutron powder diffraction of Co-Fe at $T$ = 4~K as a function of
	applied magnetic field.  The difference between the $T$ = 40~K diffractogram and the $T$ = 4~K
	diffractogram (open circles) along with profile fits to intensities (red line from model \#1 for 4 T and 1 T data in Table II) are shown for (a) 4~T, (b) 1~T, and (c) 10~mT.  Additionally, the difference between the $T$ = 4~K up-neutron-polarization diffractogram and the $T$ = 4~K down-neutron-polarization diffractogram (open circles) along with profile fits to intensity (red line from model \#5 for polarized data in Table II)  are shown for (d) 1~T.  Uncertainty bars on experimental data points are statistical in nature representing one standard deviation from the mean, using counting statistics.}
	\label{fig:KCoFeFig3}
\end{figure*}

\section{Results and Analyses}
The nuclear crystal structure of Co-Fe can be modeled with space group $Fm\overline{3}m$ (No.~225), 
where ferricyanide molecules and cobalt ions are alternately centered on the high symmetry points of 
the unit cell, with heavy-water bound to cobalt when ferricyanide is absent, and potassium ions and 
heavy-water molecules filling in voids.\cite{Herren1980,Buser1977,Hanawa2003}  This structure is used 
as a starting point to fit the $T$~=~40~K thermally quenched Co-Fe contribution to the measured 
intensity profile, Fig.~\ref{fig:KCoFeFig2}, which also has sample mount contributions due to $P63/mmc$ 
(No.~194) D$_2$O and $Fm\overline{3}m$ (No.~225) aluminum.\cite{Dowell1960,Hull1917}  Incomplete 
trapping of the high-temperature state in Co-Fe gives rise to a highly microstrained nuclear 
structure,\cite{Hanawa2003} and to account for this effect during refinement, we use an asymmetric 
double sigmoidal peak shape, namely
\begin{equation}
y_{a2s}~=~\frac{I}{2.49 w} \left(  1 - \frac{1}{1+ e^{- \frac{\theta -\theta _c}{3.43w} } } \right)    \left(   \frac{1} { 1+ e^{- \frac{\theta -\theta _c }{w} } }\right)~~~,
\end{equation}
where $I$ is the intensity, $w$ is the width, and $\theta _c$ is the center of the reflection, and 
these fits yield an effective lattice constant of 10.23 $\mathrm{\AA}$.  Observed Co-Fe reflections 
that can be clearly separated from sample holder reflections are used to extract structure factors.  
In modeling the unit cell, the cobalt to iron ratio was determined with EDS, while the room temperature 
oxidation states with FT-IR.  The carbon and nitrogen content were established with combustion analysis, and the potassium 
ions provide charge balance.  Finally, the heavy-water concentration and positions were refined along 
with the scale factor to fit the structure factors.

\begin{table*} [!htb]
\caption{Comparison of the eight magnetic models, as described in the text, numbered (\#) $1-8$ 
for Co-Fe at $T$ = 4~K in different magnetic fields tabulated as 
``cond.", which is shorthand for experimental condition, where ``$P$" designates the data acquired with 
polarized neutrons.
Here, ``align." is short
for ``moment alignment," where + denotes parallel alignment of moments and - denotes antiparallel alignment of
moments.  The units of $m_{z,Co}$ and $m_{z,Fe}$ are $\mu _B$, and the units of M are $\mu _B$~mol$^{-1}$.  The sum of the residuals are normalized to model 1 for each experimental condition.\label{table2}}
\begin{ruledtabular}
\begin{tabular}{ c c c c c c c c c c c c c }
\ \# &	cond. &	align.&	$g_{S,Co}$ &	$g_{L,Co}$ &	$g_{S,Fe}$ &	$g_{L,Fe}$ &	$J_{z,Co}$ &	$J_{z,Fe}$ &	$m_{z,Co}$ &	$m_{z,Fe}$ &	M	& $\sum_{j}{residual^2}$ \\
\hline
1 & 4~T & + & 10/3 & 1 & 2/3 & 4/3 & 0.63 $\pm$ 0.02 & 0.13 $\pm$ 0.03 & 2.7 & 0.3 & 2.6 & 1.000 \\
2 & 4~T & - & 10/3 & 1 & 2/3 & 4/3 & 0.62 $\pm$ 0.02 & 0.00 $\pm$ 0.02 & 2.7 & 0.0 & 2.4 & 1.019 \\[4pt]
3 & 4~T & + & 10/3 & 1 & 2 & 0 & 0.64 $\pm$ 0.02 & 0.12 $\pm$ 0.04 & 2.8 & 0.2 & 2.6 & 1.000 \\
4 & 4~T & - & 10/3 & 1 & 2 & 0 & 0.63 $\pm$ 0.02 & 0.00 $\pm$ 0.11 & 2.7 & 0.0 & 2.5 & 1.020 \\[4pt]
5 & 4~T & + & 2 & 0 & 2/3 & 4/3 & 1.08 $\pm$ 0.03 & 0.16 $\pm$ 0.03 & 2.2 & 0.3 & 2.1 & 1.028 \\
6 & 4~T & - & 2 & 0 & 2/3 & 4/3 & 1.04 $\pm$ 0.03 & 0.00 $\pm$ 0.03 & 2.1 & 0.0 & 1.9 & 1.049 \\[4pt]
7 & 4~T & + & 2 & 0 & 2 & 0 & 1.09 $\pm$ 0.03 & 0.12 $\pm$ 0.04 & 2.2 & 0.2 & 2.1 & 1.030 \\
8 & 4~T & - & 2 & 0 & 2 & 0 & 1.05 $\pm$ 0.03 & 0.00 $\pm$ 0.03 & 2.1 & 0.0 & 1.9 & 1.048 \\
\hline
1 & 1~T & + & 10/3 & 1 & 2/3 & 4/3 & 0.37 $\pm$ 0.03 & 0.20 $\pm$ 0.09 & 1.6 & 0.4 & 1.7 & 1.000 \\
2 & 1~T & - & 10/3 & 1 & 2/3 & 4/3 & 0.37 $\pm$ 0.02 & 0.00 $\pm$ 0.04 & 1.6 & 0.0 & 1.4 & 1.020 \\[4pt]
3 & 1~T & + & 10/3 & 1 & 2 & 0 & 0.38 $\pm$ 0.03 & 0.12 $\pm$ 0.07 & 1.6 & 0.2 & 1.6 & 1.004 \\
4 & 1~T & - & 10/3 & 1 & 2 & 0 & 0.38 $\pm$ 0.05 & 0.00 $\pm$ 0.11 & 1.6 & 0.0 & 1.5 & 1.019 \\[4pt]
5 & 1~T & + & 2 & 0 & 2/3 & 4/3 & 0.64 $\pm$ 0.04 & 0.20 $\pm$ 0.09 & 1.3 & 0.4 & 1.4 & 1.001 \\
6 & 1~T & - & 2 & 0 & 2/3 & 4/3 & 0.64 $\pm$ 0.10 & 0.00 $\pm$ 0.21 & 1.3 & 0.0 & 1.2 & 1.022 \\[4pt]
7 & 1~T & + & 2 & 0 & 2 & 0 & 0.65 $\pm$ 0.04 & 0.13 $\pm$ 0.07 & 1.3 & 0.3 & 1.3 & 1.006 \\
8 & 1~T & - & 2 & 0 & 2 & 0 & 0.63 $\pm$ 0.09 & 0.00 $\pm$ 0.08 & 1.3 & 0.0 & 1.1 & 1.021 \\
\hline
1 & $P$ & + & 10/3 & 1 & 2/3 & 4/3 & 0.38 $\pm$ 0.01 & 0.06 $\pm$ 0.02 & 1.6 & 0.1 & 1.5 & 1.000 \\
2 & $P$ & - & 10/3 & 1 & 2/3 & 4/3 & 0.40 $\pm$ 0.02 & 0.00 $\pm$ 0.01 & 1.7 & 0.0 & 1.5 & 1.058 \\[4pt]
3 & $P$ & + & 10/3 & 1 & 2 & 0 & 0.36 $\pm$ 0.02 & 0.18 $\pm$ 0.05 & 1.6 & 0.4 & 1.6 & 1.004 \\
4 & $P$ & - & 10/3 & 1 & 2 & 0 & 0.40 $\pm$ 0.02 & 0.00 $\pm$ 0.02 & 1.7 & 0.0 & 1.5 & 1.065 \\[4pt]
5 & $P$ & + & 2 & 0 & 2/3 & 4/3 & 0.68 $\pm$ 0.02 & 0.06 $\pm$ 0.02 & 1.4 & 0.1 & 1.3 & 0.995 \\
6 & $P$ & - & 2 & 0 & 2/3 & 4/3 & 0.71 $\pm$ 0.02 & 0.00 $\pm$ 0.02 & 1.4 & 0.0 & 1.3 & 1.058 \\[4pt]
7 & $P$ & + & 2 & 0 & 2 & 0 & 0.66 $\pm$ 0.03 & 0.16 $\pm$ 0.06 & 1.3 & 0.3 & 1.3 & 1.005 \\
8 & $P$ & - & 2 & 0 & 2 & 0 & 0.74 $\pm$ 0.03 & 0.00 $\pm$ 0.05 & 1.5 & 0.0 & 1.3 & 1.079 \\

\end{tabular}
\end{ruledtabular}

\end{table*}

To begin, refinement yielded interstitial 
heavy-water pseudo-atoms at the 8c position ($n$ = 0.618, $B$ = 5) and the 32f position 
(x = 0.3064, $n$ = 0.333, $B$ = 20), after which all other parameters were fixed 
(Table~\ref{table1}) and Fourier components of the heavy-water were further refined to give the 
interstitial distribution shown in Fig.~\ref{fig:KCoFeFig1}.  These refinements give a chemical 
formula of K$_{0.27}$Co[Fe(CN)$_6$]$_{0.73}$[{D$_2$O}$_6$]$_{0.27}\cdot$1.42D$_2$O. 
Moreover, at room temperature, the more complete chemical formula with oxidation states of the metal 
ions included is K$_{0.27}$Co$^{2+}_{0.94}$Co$^{3+}_{0.06}$[Fe$^{3+}$(CN)$_6$]$_{0.58}$ 
[Fe$^{2+}$(CN)$_6$]$_{0.15}$[{D$_2$O}$_6$]$_{0.27}\cdot$1.42D$_2$O, or more compactly represented 
by Co$^{2+}_{0.94}$Co$^{3+}_{0.06}$Fe$^{3+}_{0.58}$Fe$^{2+}_{0.15}$.

Having highly ionic wavefunctions, the magnetic ground states of Co and Fe in Co-Fe are well described with ligand field theory.\cite{Figgis2000}  As displayed in Fig.~\ref{fig:KCoFeFig1}, the iron atoms are octahedrally coordinated by carbon atoms that introduce a ligand field splitting parameter ($\Delta _{\mathrm{Fe}}$) of approximately 0.70~aJ (35,000 cm$^{-1}$ or 4.3 eV), and typical Fe Racah parameters put $d^5-$Fe$^{3+}$ into a $^2T_{2g}$ ground state, and $d^6-$Fe$^{2+}$ into a diamagnetic $^1A_{1g}$ ground state.  Similarly, cobalt atoms are octahedrally coordinated with oxygen and nitrogen atoms to give  $\Delta _{\mathrm{Co}^{3+}}~\approx~$0.46~aJ (23,000 cm$^{-1}$ or 2.9 eV) for $d^6-$Co$^{3+}$ that has a diamagnetic $^1A_{1g}$ ground state, and $\Delta _{\mathrm{Co}^{2+}}~\approx~$0.20~aJ (10,000 cm$^{-1}$ or 1.2 eV) for $d^7-$Co$^{2+}$ that has a $^4T_{1g}$ ground state, using typical Co Racah parameters.  At temperatures much less than the spin-orbit coupling energy, only the lowest energy total angular momentum levels are appreciably occupied, so that the relevant states are Fe$^{3+}$[$J = 1/2$, $g_J = (2+4k)/3$, $g_L = 4k/3$, $g_S = 2/3$] and Co$^{2+}$[$J = 1/2$, $g_J = (10+2Ak)/3$, $g_L = 2Ak/3$, $g_S = 10/3$], where $A$ is expected to be nearly 1.5 due to the weak ligand field, and $k$ is the orbital reduction parameter.  It is worth noting that analogous orbitally degenerate terms have been observed for $d^5-$Fe$^{3+}$($^2T_{2g}$) in K$_3$Fe(CN)$_6$,\cite{Figgis1969} and  $d^7-$Co$^{2+}$($^4T_{1g}$) in K$_{1.88}$Co[Fe(CN)$_6$]$_{0.97}\cdot$3.8H$_2$O and Na$_{1.52}$K$_{0.04}$Co[Fe(CN)$_6$]$_{0.89} \cdot$3.9H$_2$O.\cite{Matsuda2010}  Alternatively, if interaction with the lattice drastically quenches the orbital moment, spin-orbit coupling no longer splits the ground states and the magnetic parameters become Fe$^{3+}$[$J = 1/2$, $g_J = 2$, $g_L = 0$, $g_S = 2$] and Co$^{2+}$[$J = 3/2$, $g_J = 2$, $g_L = 0$, $g_S = 2$].  To estimate the relative proportion of the different oxidation states in the thermally quenched state, the effective paramagnetic moment is linearized as a function of temperature for the 300~K and 100~K states,\cite{PajerowskiPHD} and the measured lattice constant is compared to a weighted average of quenched and ground-state lattice constants,\cite{Chong2011} to self-consistently give Co$^{2+}_{0.90}$Co$^{3+}_{0.10}$Fe$^{3+}_{0.54}$Fe$^{2+}_{0.19}$ for the magnetic quenched state at 100~K and below that is analyzed in detail herein.

Cooling the sample further, subsequent to quenching, the bulk magnetization measured in 10~mT showed the well-documented upturn at around 15 K corresponding to the onset of magnetic order.  Therefore, additional NPD was performed at 4~K in applied fields of 10~mT, 1~T, and 4~T, Fig.~\ref{fig:KCoFeFig3}, to compare to the scattering in the paramagnetic state.  Furthermore, polarized NPD was performed at 4~K in an applied field of 1~T, Fig.~\ref{fig:KCoFeFig3}~(d), where the difference between diffractograms for up and down incident neutron polarizations increases signal to noise of the measured magnetic structure at reflections with large nuclear contributions.  
For each of the three experimental conditions where magnetic scattering is observed, we compare the results of 
eight plausible but different models that all have moments along the applied field.  Specifically, 
each possible case considers various combinations of the parallel or antiparallel alignment of Co and Fe 
moments when each ion possesses either spin-only or orbitally degenerate magnetic states, Table~\ref{table2}. 
The analyses indicate that most magnetism resides on the Co 4a site for all models, with a parallel alignment of Fe and Co moments giving the best fits and  $\chi ^2$ surfaces suggesting a reduced but present orbital moment on both ions.  No magnetic scattering is observed in 10~mT, and increased coherent magnetic scattering appears with increasing field, which is consistent with the presence of significant random anisotropy, where a correlated spin glass (CSG) is the ground state and sufficiently large fields cause entrance into a 
ferromagnetic phase with wandering axis (FWA) state or at even larger fields a nearly collinear (NC) state.\cite{Chudnovsky1986}

\begin{figure} [t!]
	\includegraphics[width=87mm]{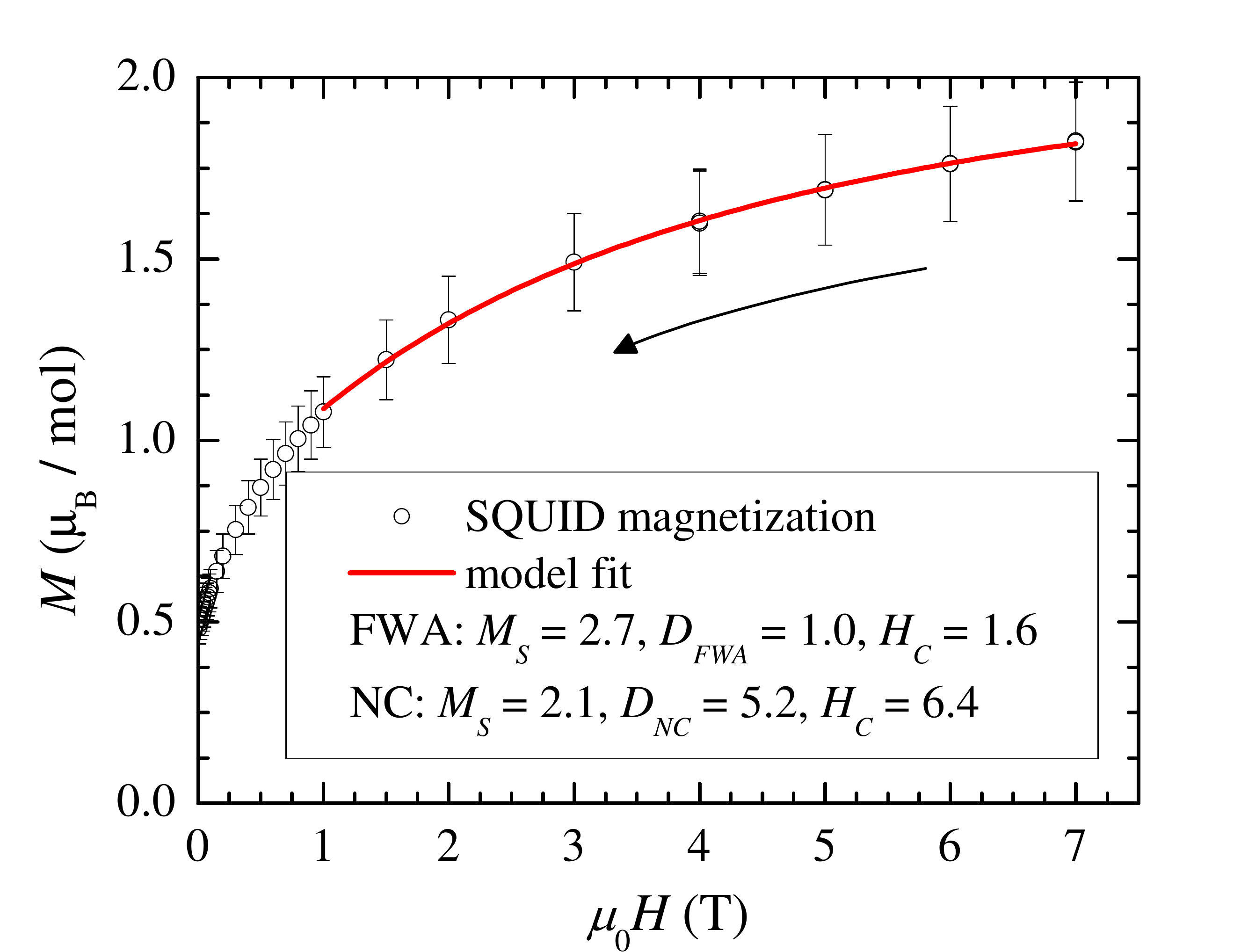}
	\caption{(Color online) SQUID  magnetization of Co-Fe at $T$~=~2~K.  Experimental magnetization is shown as open circles (SQUID magnetization) and model fits as a red line (model fit), where FWA and NC are visually indistinguishable; $M (1~\mathrm{T}) = 1.1 \pm 0.1~\mu _B$~mol$^{-1}$ and $M (4~\mathrm{T}) = 1.6 \pm 0.2~\mu _B$~mol$^{-1}$.  The units of $M_S$ are $\mu_ B$~mol$^{-1}$, and $H_C$, $D_{FWA}$, and $D_{NC}$ are all units of Tesla.  Uncertainty bars represent one standard deviation from the mean, where statistics are generated by measuring the magnetization of the 14 synthesis batches required to generate 4.37~g for NPD. }
	\label{fig:KCoFeFig4}
\end{figure}

Analytical expressions for the magnetization process for magnets with random anisotropy are available\cite{Chudnovsky1986} for the Hamiltonian
\begin{eqnarray}
\mathcal{H}~=~ -\mathcal{J} \sum_{i,j}{S_i \cdot S_j}  -  D_r \sum_{i}{(\hat{n_i} \cdot S_i )^2}  \nonumber \\
   -D_c \sum_{i}{(S_i ^z )^2 }  - g \mu _B \sum_{i} {H \cdot S_i}~~~,
\end{eqnarray}
where $\mathcal{J}$ is the superexchange constant, $S$ is the spin operator, $D_r$ is strength of the random anisotropy, $\hat{n}$ is the direction of the random on-site anisotropy, $D_c$ is the strength of the coherent anisotropy, $g$ is the Land$\acute{\mathrm{e}}$ factor, and $H$ is the applied field.  In the FWA regime where the applied field energy is larger than the random anisotropy energy but much less than the exchange field, the low temperature magnetization is
\begin{eqnarray}
M_{FWA}&=& M_S   -   \frac{6 \sqrt{2}  D_r^2 \Omega  M_S } {5 \pi ^2 a^3  (z \mathcal{J})^{3/2 } (H +  H_C  )^{1/2}  } \nonumber \\
 &=&   M_S  \left(1 -   \frac{D_{FWA}^{1/2 }}  {(H +  H_C  )^{1/2} }\right)~~~,
\end{eqnarray}
where $M_S$ is the saturation magnetization, $\Omega$ is the integrated local anisotropy correlation function, $z$ is the number of magnetic neighbors, $a$ is the mean distance between neighboring spin sites, $H_C$ is the coherent anisotropy field, and $D_{FWA}$ is a measure of the random anisotropy to superexchange strengths.  For the NC phase that is reached when the applied field and coherent anisotropy field are much larger than the exchange field, the low temperature magnetization is
\begin{eqnarray}
M_{NC}   &=&  M_S   -   \frac{4  D_r^2  M_S } {15 a^6  (H +  H_C  )^2  }  \nonumber \\
&=&   M_S  \left(\frac{1 -   D_{NC} ^2} {(H +  H_C  )^2  }\right)~~~,
\end{eqnarray}
where $D_{NC}$ is a measure of the random anisotropy.  Based upon the magnetic ordering temperature, Co-Fe is expected to be in an FWA-like phase, and although both FWA and NC expressions may be fit to the low temperature magnetization data, Fig.~\ref{fig:KCoFeFig4}, the parameters extracted by the NC fit are not consistent with the derivation limit for Eq. 11, further suggesting an FWA-like state.

\begin{figure} [t!]
	\includegraphics[width=87mm]{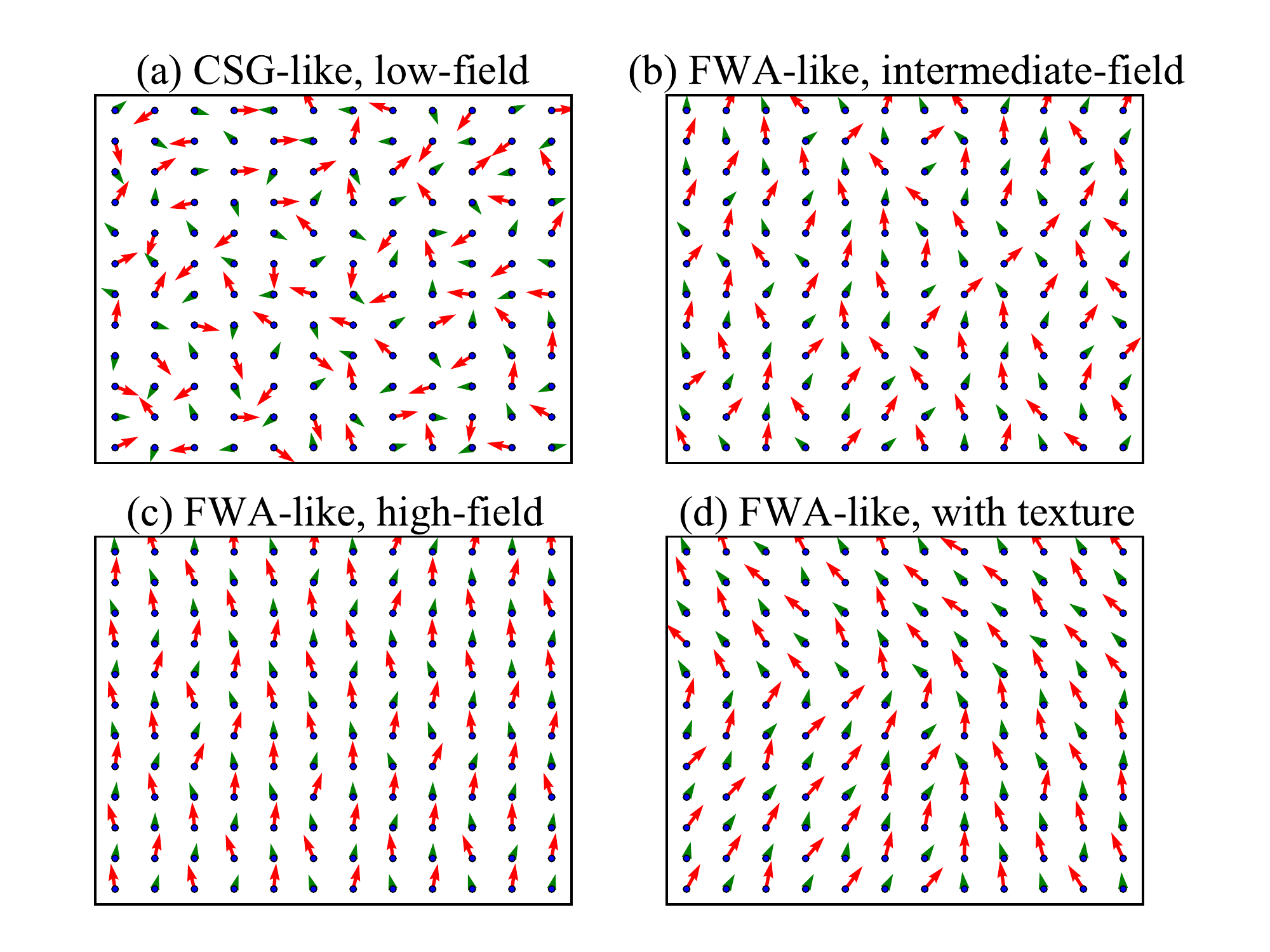}
	\caption{(Color online) An illustration of magnetic structure for different magnetic field regimes.  Here, the magnetic field points towards the top of the page, short arrows represent iron moments, and long arrows represent cobalt moments.  (a) The Co-Fe sample cooled in zero field has a CSG-like state with no average on-site moment, as shown for the measurement of magnetic scattering in 10 mT, Fig. 3 (c).  (b) The application of magnetic field cants the moments towards the field (Fig. 3 (b) and (d) ), and (c) larger fields induce larger average moments (Fig. 3 (a) ).  (d) More complicated mesoscopic states that contain texture are also possible, but are not unambiguously determined with our data. }
	\label{fig:KCoFeFig5}
\end{figure}

\section{Discussion}

\begin{figure}[t!]
	\includegraphics[width=87mm]{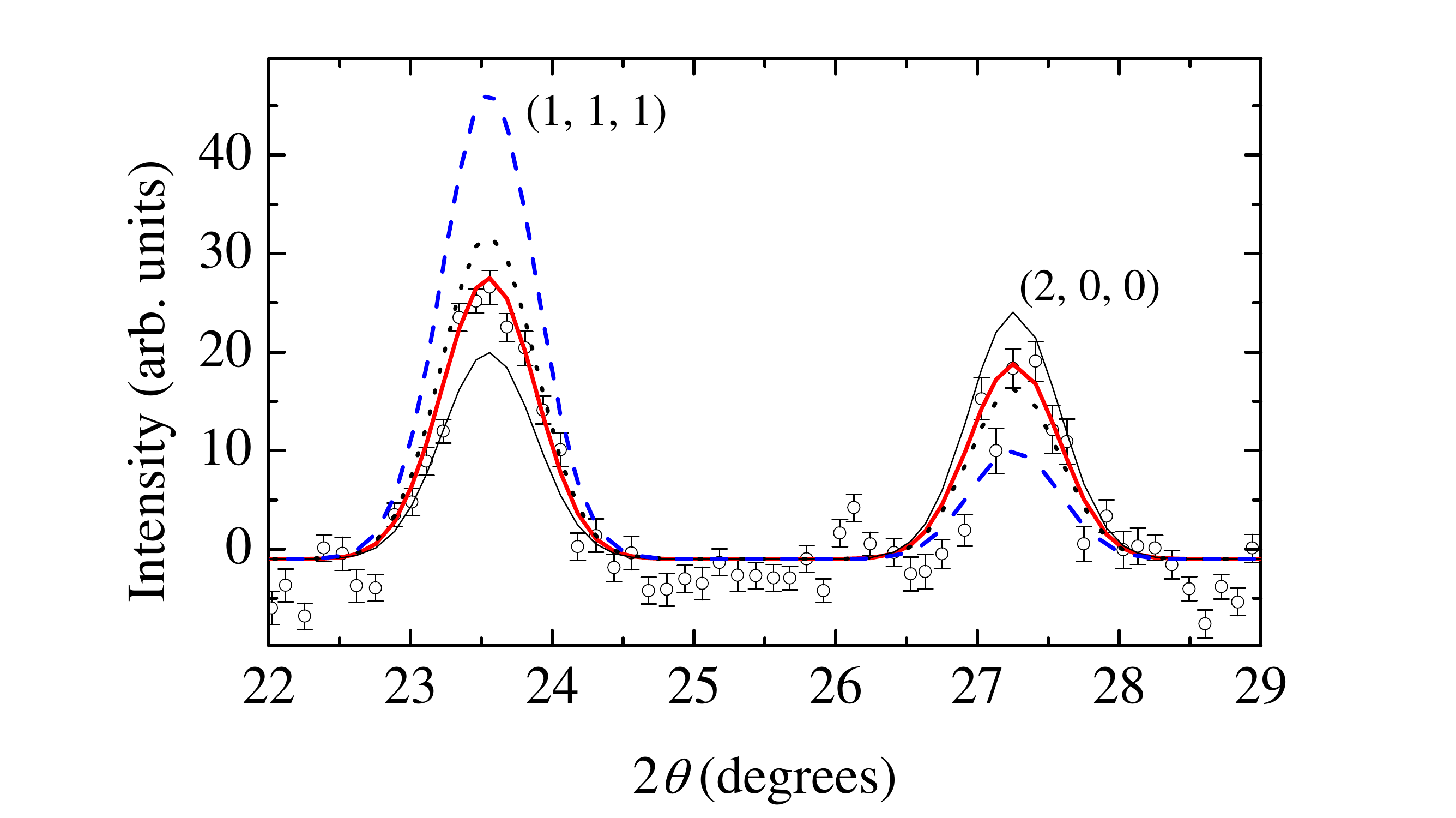}
	\caption{(Color online) Visual comparison of magnetic configurations for Co-Fe.  Here, the low angle H = 4 T data (open circles) are set side by side with model \#1 for 4 T from Table II (thick-solid, red line), model \#2 for 4 T from Table II (dotted, black line), a ferrimagnetic structure with a 3:1 Co:Fe spin ratio (dashed, blue line), and a ferromagnetic structure with a 3:1 Co:Fe spin ratio (thin-solid, black line).  Uncertainty bars are representative of one standard deviation from the mean, using counting statistics. }
	\label{fig:KCoFeFig6}
\end{figure}

We have presented neutron diffraction and bulk magnetization measurements of K$_{0.27}$Co[Fe(CN)$_6$]$_{0.73}$[{D$_2$O}$_6$]$_{0.27}\cdot$1.42D$_2$O that suggest a CSG ground state that enters a FWA-like state in applied magnetic fields of the order 1~T and larger (Fig.~\ref{fig:KCoFeFig5}).  This conclusion is based upon the field dependence of the magnetization, and particularly the diffraction experiments that show an absence of long range order in the 10~mT data and ordered moments that are induced by the applied field, ruling out a high-field domain magnetization process.  This random-anisotropy-based magnetization process explains the appreciable slope observed for Co-Fe even in fields of 7 T at 2 K (Fig. 4), in a way similar to the magnetization process at low temperature and high magnetic fields for the vanadium tetracyanoethylene molecular magnet prepared using solvent based methods.\cite{Zhou1993}  This magnetization process is different than for other reported cubic complex cyanide systems that have magnetically ordered ground states and saturate magnetization at 2 K and 7 T.\cite{Mihalik2010,Kumar2004,Kumar2005}  The CSG ground state is consistent with previous AC-susceptibility measurements of K$_{1-2x}$Co$_{1+x}$[Fe(CN)$_6$]$\cdot y$H$_2$O (0.2~$\leq~x~\leq$~0.4, $y~\approx$~5) that showed glassy behavior,\cite{Pejakovic2002} although the relative orientation we find for Co and Fe at 1~T is contrary to the XMCD experiment that reported antiparallel Co and Fe at 1~T in Rb$_{1.8}$Co$_4$[Fe(CN)$_6$]$_{3.3}\cdot$13H$_2$O and K$_{0.1}$Co$_4$[Fe(CN)$_6$]$_{2.7}\cdot$18H$_2$O.\cite{Champion2001}  For fields of 1~T and 4~T, we find a parallel alignment of Co and Fe moments minimizes the residuals between model and data, and this alignment is clearly seen for the low-angle 4~T peaks, Fig.~\ref{fig:KCoFeFig6}.  However, the lack of coherent scattering for CSG means we cannot strictly discuss the nature of the superexchange interaction in the ground state, although we infer some significant ferromagnetic character based upon the high-field regime.  One must also be careful about applying qualitative Goodenough-Kanamori rules to this system,\cite{Goodenough1955, Goodenough1958, Kanamori1959} as the single-electron states are not only mixed from electrostatic interactions, but also due to the aforementioned presence of spin-orbit coupling.  A band structure calculation using a full potential linearized augmented plane wave method resulted in an antiferromagnetic ground state for Co-Fe, with +0.296~$\mu _B$ on the Co site and -0.280~$\mu _B$ on the Fe site,\cite{takegahara2002} although such values do not agree with experimental findings.

For the Co-Fe presented, care was taken to ensure that the average particle size was greater than $\approx$100~nm (as measured by TEM) to avoid finite size effects,\cite{Pajerowski2007} and the FWA-like phase can explain the previously reported changes in low temperature high field magnetization with particle size as a tuning of the local random anisotropy with size, larger particles requiring higher fields to saturate at base-temperature as they are deeper in a glassy phase.  The saturation value for bulk K$_{0.27}$Co[Fe(CN)$_6$]$_{0.73}$[{D$_2$O}$_6$]$_{0.27}\cdot1.42$D$_2$O, $M_S = 2.7 \pm 0.3~\mu _B$~mol$^{-1}$, is comparable to a variety of similar states with different degrees of orbital reduction on Co and Fe sites; $exempli~gratia$, considering complete orbital moments on both Co and Fe gives $M_S \approx 2.5~\mu _B$~mol$^{-1}$ and spin-only moments gives $M_S \approx 3.2~\mu _B$~mol$^{-1}$.

The NPD experiments show a model-independent increase in coherent magnetic scattering as a function of field, and a slightly form-factor dependent ratio of Co to Fe moments, Table~\ref{table2}.  A parallel alignment of Co and Fe is found for both 1~T and 4~T, and when an antiparallel alignment is forced, Fe moments go to zero to achieve best fits.  The moment ratio is heavily dictated by the low-Q peaks where the form-factor has little effect, while the scale of the moments is different depending upon the presumed shape of the scatterer.  Previous neutron diffraction measurements have shown covalency effects, due to  $\sigma$-bonding and  $\pi$-back-bonding with CN, to be important in the chemically similar CsK$_2$[Fe(CN)$_6$].\cite{Figgis1990}  Covalency can increase the direct-space size of the moments, thereby decreasing the reciprocal-space size even in the presence of orbital moments.  For Co-Fe, the smaller reciprocal-space form-factors give magnetizations most similar to those determined by SQUID, although covalency makes assignment of orbital and spin magnetism based upon spatial distribution inconclusive.  We do not refine the form-factor in this manuscript because high parameter covariance is introduced.  Finally, small quantitative differences between SQUID and NPD moment values may also be due to sample inhomogeneity overestimating the NPD moments, and unitemized experimental uncertainties due to the complicated and highly un-stoichiometric formulation of the Co-Fe material, but our conclusions remain robust with respect to such perturbations.  The line-widths for magnetic and nuclear NPD are similar, suggesting comparable domain sizes for the scattering objects, but it is possible that the induced moments have a texture over some other length scale, Fig. 5 (d), a possibility suggested by cluster-glass behavior in AC susceptibility studies.\cite{Pejakovic2002}  As shown in the Appendix, regions of coherent magnetization at an angle $\xi$ from the applied field would rescale the measured NPD longitudinal moment by $\cos\xi \sqrt{\frac{2}{\cos^2\xi + 1}}$, but the similarity between the polarized and unpolarized magnetic diffraction further suggests that such an effect is small in our samples.

\section{Conclusions}
The neutron diffraction and bulk magnetization measurements of Co-Fe suggest a magnetization process that evolves from a correlated spin glass to a quasi-ferromagnetic state with increasing magnetic field, where average Co and Fe moments are induced to lie along the applied field.  When considering memory storage applications of molecule based magnetic materials, structure-property relationships that may give rise to coherent and random anisotropy will be important to consider.

\begin{acknowledgments}
DMP acknowledges support from the NRC/NIST post-doctoral associateship program. Research at High Flux Isotope Reactor at ORNL was sponsored by the Scientific User Facilities Division, Office of Basic Energy Sciences, U. S. Department of Energy.  This work was supported, in part, by NSERC, CFI, and NSF through Grants No.~DMR-1005581 (DRT) and No.~DMR-0701400 (MWM).
\end{acknowledgments}

\appendix*
\section{Effect of Canting on Intensity}
A powder sample consisting of domains canted at an angle $\xi$ away from the applied field, with random rotational distribution, may give the same unpolarized NPD signal as domains along the field, but with a different magnetic moment.  With the magnetic field along the z-axis, and the scattering vector along the x-axis, the uncanted magnetic moment is simply
\begin{equation}
\mathbf{M_u}~=~(0,0,M)~~~,
\end{equation}
and the canted moment can be expressed as
\begin{equation}
\mathbf{M_{canted}}~=~M(\sin\xi\cos\phi,\sin\xi\sin\phi,\cos\xi)~~~,
\end{equation}
where $\xi$ is the canting angle, $\phi$ is the rotation angle about the field, and $M$ is the magnitude of the magnetic moment.  The interaction vector\cite{Schweizer2006}
\begin{equation}
\left|\mathbf{M_{\bot}}\right|^2~=~\sum_{\alpha,\beta}{\left(\delta_{\alpha\beta}-\widehat{Q}_{\alpha}\widehat{Q}_{\beta}\right)M^*_{\alpha}M_{\beta}}~~~,
\end{equation}
then gives a dependence of the intensity on the canting angle such that
\begin{equation}
\left|\mathbf{M_{u,\bot}}\right|^2~=~M^2~~~,
\end{equation}
and
\begin{equation}
\left|\mathbf{M_{canted,\bot}}\right|^2~=~M^2 {\frac{\cos^2\xi + 1}{2}}~~~,
\end{equation}
where random $\phi$-angles have been averaged over.  Therefore, an uncanted model with a z-component (relevant to compare with longitudinal magnetization) of $M$ can give the same unpolarized NPD intensity as a canted model with a z-component of $M \cos\xi \sqrt{\frac{2}{\cos^2\xi + 1}}$.


%


\begin{thebibliography}{22}
\bibitem []{NISTdisclaimer}Certain commercial equipment, instruments, or materials are identified in this paper to foster understanding. Such identification does not imply recommendation or 
endorsement by the National Institute of Standards and Technology, nor does it imply that the materials or equipment identified are necessarily the best available for the 
purpose.%
\end{thebibliography}
\end{document}